\documentclass[10pt,conference,a4paper]{IEEEtran}
\usepackage[T1]{fontenc}% optional T1 font encoding
\usepackage{amsmath}
\usepackage[cmintegrals]{newtxmath}
\usepackage{url}
\usepackage{soul,color,graphicx,float,epstopdf}
\usepackage{tablefootnote,textcomp,gensymb}
\usepackage{eurosym,booktabs,multirow}
\usepackage[caption=false]{subfig}
\usepackage{amsfonts} 
\usepackage{algorithmic}
\usepackage{array}
\usepackage{stfloats}
\usepackage{float}
\usepackage{mathtools}
\usepackage{graphicx}
\usepackage{pdfpages}
\usepackage{subfig} 
\usepackage{balance}
\usepackage{xcolor}
\usepackage{comment}
\usepackage[normalem]{ulem}
\usepackage{dsfont}
\usepackage[linesnumbered, ruled]{algorithm2e}
\usepackage{enumitem}
\usepackage{tikz}
\usepackage{cite}

\usepackage{geometry}
\usetikzlibrary{arrows,shapes,positioning,shadows,trees}

\tikzset{
  basic/.style  = {draw, text width=4cm, drop shadow, font=\sffamily, rectangle},
  root/.style   = {basic, rounded corners=2pt, thin, align=center,
                   fill=red!20},
  level 2/.style = {basic, rounded corners=6pt, thin,align=center, fill=blue!20,
                   text width=9em},
  level 3/.style = {basic, thin, align=left, fill=pink!20, text width=8em}
}

\definecolor{Orange}{rgb}{1,0.5,0}
\newcommand{\eqr}[1]{$#1$}

\newcommand*{\rom}[1]{\expandafter\@slowromancap\romannumeral #1@}

\newcommand{\todo}[1]{\textsf{\textbf{\textcolor{Orange}{[#1]}}}}

\begin{document}
% \title{Reconfigurable Intelligent Surface Multi-reflection Enhancement: A Deep Reinforcement Learning Based Method}
\title{A DRL-based Reflection Enhancement Method for RIS-assisted Multi-receiver Communications}

\author{\IEEEauthorblockN{
Wei Wang\IEEEauthorrefmark{1},
Peizheng Li\IEEEauthorrefmark{2},
Angela Doufexi\IEEEauthorrefmark{1},
Mark A Beach\IEEEauthorrefmark{1}
}\\ 
\IEEEauthorblockA{\IEEEauthorrefmark{1} University of Bristol, UK;
\IEEEauthorrefmark{2} Bristol Research and Innovation Laboratory, Toshiba Europe Ltd., UK\\
Email: {\{wei.wang, A.Doufexi, M.A.Beach\}@bristol.ac.uk;}
peizheng.li@toshiba-bril.com}}
\maketitle

\begin{abstract}
In reconfigurable intelligent surface (RIS)-assisted wireless communication systems, the pointing accuracy and intensity of reflections depend crucially on the 'profile,' representing the amplitude/phase state information of all elements in a RIS array. The superposition of multiple single-reflection profiles enables multi-reflection for distributed users. However, the optimization challenges from periodic element arrangements in single-reflection and multi-reflection profiles are understudied. The combination of periodical single-reflection profiles leads to amplitude/phase counteractions, affecting the performance of each reflection beam. This paper focuses on a dual-reflection optimization scenario and investigates the far-field performance deterioration caused by the misalignment of overlapped profiles. To address this issue, we introduce a novel deep reinforcement learning (DRL)-based optimization method. Comparative experiments against random and exhaustive searches demonstrate that our proposed DRL method outperforms both alternatives, achieving the shortest optimization time. Remarkably, our approach achieves a $1.2$ dB gain in the reflection peak gain and a broader beam without any hardware modifications.
\end{abstract}

\section{Introduction}
\label{sec:intro}
Higher frequency bands, such as the millimeter wave (mm-Wave) and terahertz wave (THz-Wave) have attracted attention due to the abundant spectrum resource availability, potential higher data rate, and lower latency. However, mm-Wave and THz-Wave also suffer the issues of severe propagation loss, short propagation distance, and high sensitivity to blockage. reconfigurable intelligent surface (RIS) is regarded as a promising solution to address these issues~\cite{huang2019reconfigurable}. RIS evolved from the metasurface which consists of many reflecting elements.
It serves in the wireless channel, deployed between base stations (BS) and user equipment (UE). By establishing two additional line of sight (LoS) channels (BS-RIS and RIS-UE), RIS will provide a better communication quality than the initial non-line of sight (NLoS) channel (BS-UE) caused by obstacles.
Compared with the traditional methods targeting the improvements merely on the BS and/or UE sides, RIS unprecedentedly offers the opportunity to adjust the propagating environment directly and to improve the signal's quality.

An elementary RIS system contains a BS, UEs, and a controller. As a promising technique aiming for large coverage with low cost, the elements of RIS are mostly passive without any radio frequency (RF) chain. This critical difference sets it apart from relay schemes with similar functions. Additionally, the RIS functions as a large reflector array, compensating for its passivity and, as a result, outperforming relay schemes~\cite{bjornson2019intelligent}. The passive reflecting element can be consisted of the mounting PIN diodes~\cite{wang20201}, varactor diodes~\cite{dai2019wireless}, or radio frequency microelectromechanical systems (RF-MEMS)~\cite{debogovic2014low} on the radiation patches, and also can be realized by using liquid crystal~\cite{zhang2010voltage}. By configuring these devices or materials to operate at different states, the reflective amplitude or phase on each RIS element can be modulated, enabling the entire RIS array effectively reflect the incident wave in desired directions. The modulated amplitude or/and phase states of all the elements in one RIS array are termed as the reflection '\textbf{\textit{profile}}' in this paper.

The study of RIS has primarily focused on the design of reconfigurable reflectarrays~\cite{kamoda201160}~\cite{yang20171600} and channel estimation~\cite{hu2021two}~\cite{wang2020compressed}. However, limited attention has been given to the arrangement of reflection profiles. Some recent works, such as~\cite{bao2019multi} and~\cite{taghvaee2022enabling}, proposed novel metasurfaces that allow modulation of both phase and amplitude, presenting a theoretical basis to achieve various states of multiple reflected beams. Additionally, authors in~\cite{taghvaee2022enabling} compared the performance of different multiple-beam forming strategies, indicating that the phase-only reconfiguration can achieve a moderate side lobe level with lower complexity, power consumption, and latency compared with the amplitude and phase joint reconfiguration, space division multiplexing, and time division multiplexing. However, both~\cite{bao2019multi} and~\cite{taghvaee2022enabling} approaches establish multiple reflections from the direct superposition of single reflective profiles. In cases where single-reflection profiles, especially in a large array, exhibit a periodical distribution with the phases varying from $0$ to $360$ degrees, such direct superposition cannot guarantee RIS the optimal performance serving distributed UEs.
The intensity of each specific reflection will be disturbed by the overlapped multi-reflection profile. Therefore, investigation of its optimization methods is needed, to fully exploit the potential of RIS in supporting multi-receiver communications.

Deep reinforcement learning (DRL) has demonstrated remarkable success in solving real-world problems, including optimizing RIS and beamforming~\cite{huang2021multi}~\cite{huang2020reconfigurable}. In this paper, we employ DRL to explore the optimal superposition location of the multiple single-reflective sub-profiles, ensuring that the resulting multiple-beam reflective profile achieves its peak directivity in a given pair of directions. The DRL-based method outperforms random search significantly. Furthermore, when compared with the exhaustive search, the DRL-based approach achieves similar performance and accuracy while requiring less searching time. 
The primary contributions of this paper are summarised below:
\begin{itemize}
    \item To the best of the authors' knowledge, this is the first work that considers optimization regarding the misaligned superposition of reflective profiles for multiple-receiver applications.
    \item This study represents the first endeavor to adopt DRL for optimizing reflection profile superposition and far-field performance in RIS-assisted multiple-receiver communication scenarios.
    \item The simulation results show that the peak far-field gain has a $1.2$ dB improvement and a better coverage by simply optimizing the superposition of several sub-profiles, without any changes to the hardware.
\end{itemize}

\section{System Model and Problem Formulation}
\label{System Model and Problem Formulation}
\subsection{System Model}
\label{System Model}
We consider the downlink of a RIS-assisted mmWave system (at 28~GHz) for broadcasting applications, with a weak direct channel (shaded by obstacles) between BS and UEs. Without loss of generality, it is supposed that the RIS is a planar $30 \times 30$ reflectarray with \textit{M} elements, serving for a planar array BS with \textit{N} antennas and \textit{K} single-antenna UEs. The distance between the center point of adjacent elements is 3~mm which is around 1/4~wavelength at 28~GHz. Then, the BS-RIS and RIS-UEs channel can be denoted as $\boldsymbol{\rm{G}} \in \mathbb{C}^{M\times N}$ and $\boldsymbol{\rm{h}}_{r,k} \in \mathbb{C}^{M}$ respectively. Upon impinging on the RIS array, the incident signal undergoes phase shift and amplitude attenuation by each reflective element. Thus, the functionality of the RIS can be mathematically expressed as:
\begin{equation}
    \boldsymbol{\Theta} = [\beta_{1}e^{j\theta^{*}_{1}},\beta_{2}e^{j\theta^{*}_{2}},...,\beta_{M}e^{j\theta^{*}_{M}}]^{T}\,,
\end{equation}
where $\beta_m \in [0,1]$ is the amplitude response and $\theta^{*}_m \in [0,2\pi]$ is the phase response at the \textit{m}-th element of RIS. The received signal $y_{k}\in \mathbb{C}$ at \textit{k}-th UE can be expressed as:
\begin{equation}
    y_{k} = (\boldsymbol{\rm{h}}_{r,k} \rm{diag}(\boldsymbol{\Theta})\boldsymbol{\rm{G}}) \boldsymbol{\rm{s}}^{\rm{T}} + \it{w}_{\emph{k}}\,,
\end{equation}
where $\textbf{s} \in \mathbb{C}^{N}$ is the original signal from BS, $\it{w}_{k} \in \mathbb{C}$ is the received noise with zero mean and variance $\sigma^{2}$ at \textit{k}-th UE.

\begin{figure}[t]   
    \subfloat[\label{fig:direction_1}]{
      \begin{minipage}[t]{0.3\linewidth}
        \centering 
        \includegraphics[width=1.4in]{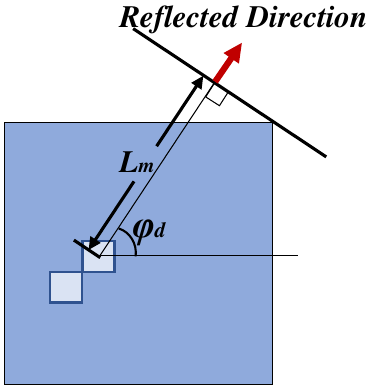}   
      \end{minipage}%
      }
      \hfill
        \subfloat[\label{fig:direction_2}]{
      \begin{minipage}[t]{0.55\linewidth}   
        \centering   
        \includegraphics[width=1.9in]{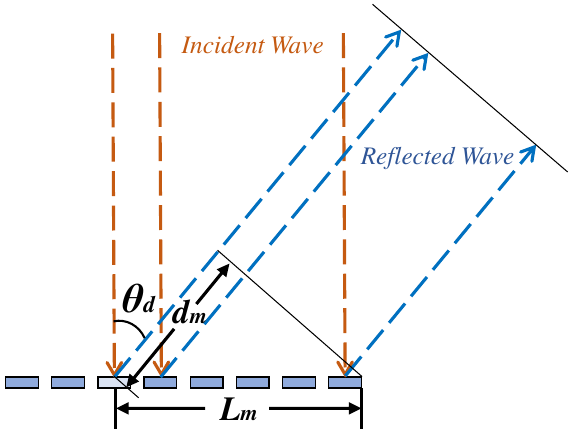}   
      \end{minipage} 
      }
      \caption{The designed arranging method of RIS array on (a) vertical view and (b) side view.
      } \label{fig:direction}
      \vspace{-0.5cm}
\end{figure}

\begin{figure}[t]   
    \subfloat[\label{fig:original_1}]{
      \begin{minipage}[t]{0.45\linewidth}
        \centering 
        \includegraphics[width=1.7in]{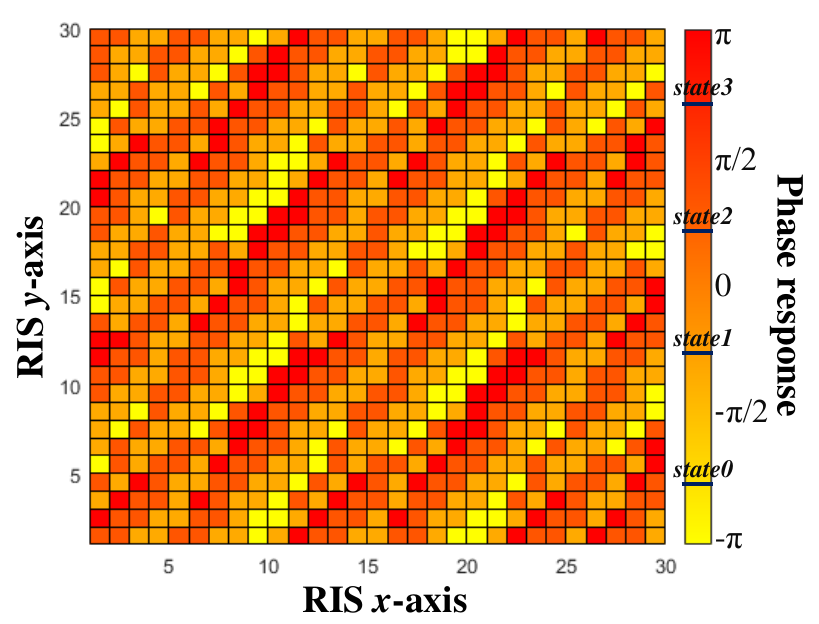}   
      \end{minipage}%
      }
      \hfill
        \subfloat[\label{fig:original_2}]{
      \begin{minipage}[t]{0.55\linewidth}   
        \centering   
        \includegraphics[width=1.7in]{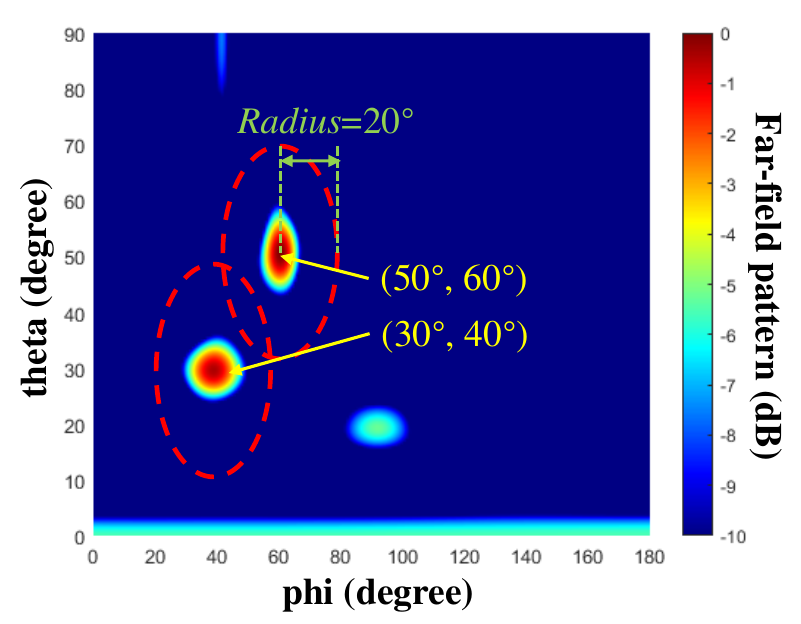}   
      \end{minipage} 
      }
      \caption{Diagrams of the reflect directions (30°, 40°) and (50°, 60°), (a) reflective profile, and (b) far-field pattern.
      } \label{fig:original}
      \vspace{-0.2cm}
\end{figure} 

\subsection{Phase Modulations at RIS}
\label{Phase Modulations at RIS}
Due to the superiority of phase-only modulation~\cite{pei2021ris} over amplitude-only modulation~\cite{cao2021design} in performance and the reduced complexity in design and fabrication compared to amplitude-phase-joint arrays~\cite{wang2022broadband}, we employ isotropic antennas with fixed amplitude response $\beta_M$ ($0.7$ as an empirical value~\cite{pei2021ris, araghi2022reconfigurable}). The mmWave channel has low multipath components and most energy is concentrated on the LoS path. So for a far-field application, the phase shift needed on each RIS element can be calculated from the optical path difference directly. As shown in Fig.~\ref{fig:direction}, for a given reflected direction [$\theta_{d}, \varphi_{d}$] under a spherical coordinate system, the phase difference $\Delta_{m}$ of \textit{m}-th RIS element can be denoted as:
\begin{equation}
    d_{m}=L_{m}\cdot \rm{sin}\theta_{d}\,,
    \label{eq:3}
\end{equation}
\begin{equation}
    \Delta_{m}=\frac{2\pi}{\lambda}\cdot d_{m}\,,
    \label{eq:4}
\end{equation}
where $L_{m}$ represents the distance from the RIS's edge to the \textit{m}-th RIS element along the reflected direction, $d_{m}$ denotes the optical path difference calculated from $L_{m}$, and $\lambda$ indicates the wavelength of the incident signals in the vacuum. Then, the phase difference of each RIS element can be calculated.

For this reason that the actually achievable phase shift $\theta^{*}$ of each RIS element cannot exceed $2\pi$ and should equal several fixed values based on its resolution, the phase difference $\Delta_{m}$ should be processed with modulo arithmetic divided by $2\pi$ and the periodicity should be introduced into the phase shift arrangement of RIS. Then, $\Delta_{m}$ can be transferred to the real phase shift $\theta^{*}$ on each element after rounding off. To ensure the symmetry of phase shift values, the phase shift used in the simulation has to offset the real phase shift value $\theta^{*}$ from $[0, 2\pi]$ to $[-\pi, +\pi]$ as shown in Fig.~\ref{fig:original_1}. In this figure, the group of $[-135^{\circ}, -45^{\circ}, 45^{\circ}, 135^{\circ}]$ has been selected as the four quantization phase states to make the RIS reflectarray having a 2-bit resolution as an example.

\subsection{Far-field Calculations of the Reflected Waves}
In calculating the RIS far-field radiation pattern, we use $\textit{OPD}_{(x,y),(i,j)}$ to denote the optical path difference between the first RIS element and the RIS element at [\textit{x},\textit{y}] on the direction [\textit{i},\textit{j}]. Then, the array factor $AF_{(i,j)}$ at the direction [\textit{i},\textit{j}] can be expressed as:
\begin{equation}
    \textit{OPD}_{(x,y),(i,j)}=D_{x}\cdot \rm{cos}(\varphi_{\textit{f j}})\rm{sin}(\theta_{\textit{f i}})+\textit{D}_{\textit{y}}\cdot \rm{sin}(\varphi_{\textit{f j}})\rm{sin}(\theta_{\textit{f i}})\,,
    \label{eq:5}
\end{equation}
\begin{equation}
    \textit{AF}_{(i,j)}=\sum_{x=0}^{X-1} \sum_{y=0}^{Y-1} I_{(x,y)}\cdot e^{j[k\cdot OPD_{(x,y),(i,j)}+\theta^{*}_{(x,y)}]}\,,
    \label{eq:6}
\end{equation}
where $D_x$ and $D_y$ represent the distance between the first RIS element and the element at [\textit{x},\textit{y}] in the x- and y-directions respectively, [$\theta_{fi}, \varphi_{fj}$] is the far-field direction, \textit{k} indicates the wavenumber which can be calculated by $k=2\pi/\lambda$, and $I_{(x,y)}$ denotes the excitation coefficient of the element at [\textit{x},\textit{y}] which is fixed to be $0.7$ here as an experienced value. Then, the far-field pattern of this RIS reflectarray can be denoted as:
\begin{equation}
    \rm{\textbf{E}}=\rm{\textbf{E}}_{single} \cdot \textbf{AF}\,,
    \label{eq:7}
\end{equation}
where $\rm{\textbf{E}}_{single}$ is the far-field pattern of the single RIS element and is set to be the ideal isotropic antenna for simplification. The far-field pattern of the direction pair $[30^{\circ}, 40^{\circ}]$ and $[50^{\circ}, 60^{\circ}]$ is exemplified in Fig.~\ref{fig:original_2},
calculated with the above equations (equation~(\ref{eq:3}) - equation~(\ref{eq:7})).

\subsection{Problem Formulation} \label{subsec:Problem Formulation}
\begin{figure*}[t]   
     \subfloat[\label{fig:schematic}]{
      \begin{minipage}[t]{0.315\linewidth}   
        \centering   
        \includegraphics[width=1.7in]{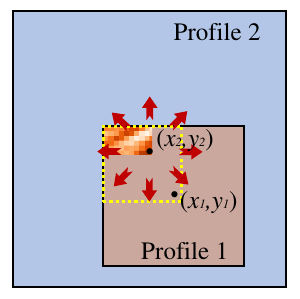}   
      \end{minipage}  
      }
    \subfloat[\label{fig:exhausted_1}]{
      \begin{minipage}[t]{0.31\linewidth}   
        \centering   
        \includegraphics[width=2.2in]{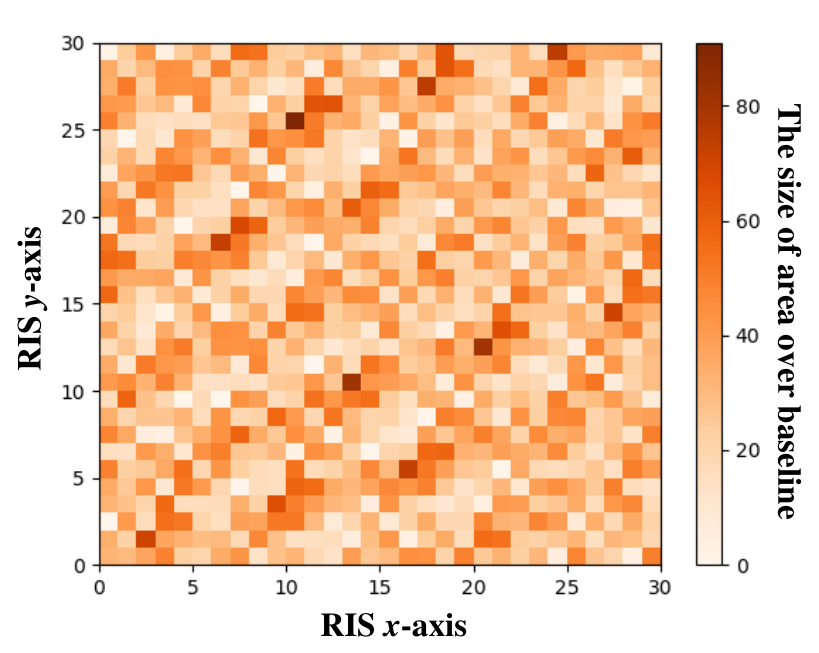}   
      \end{minipage}  
      }
     \subfloat[\label{fig:exhausted_2}]{
      \begin{minipage}[t]{0.31\linewidth}   
        \centering   
        \includegraphics[width=2.2in]{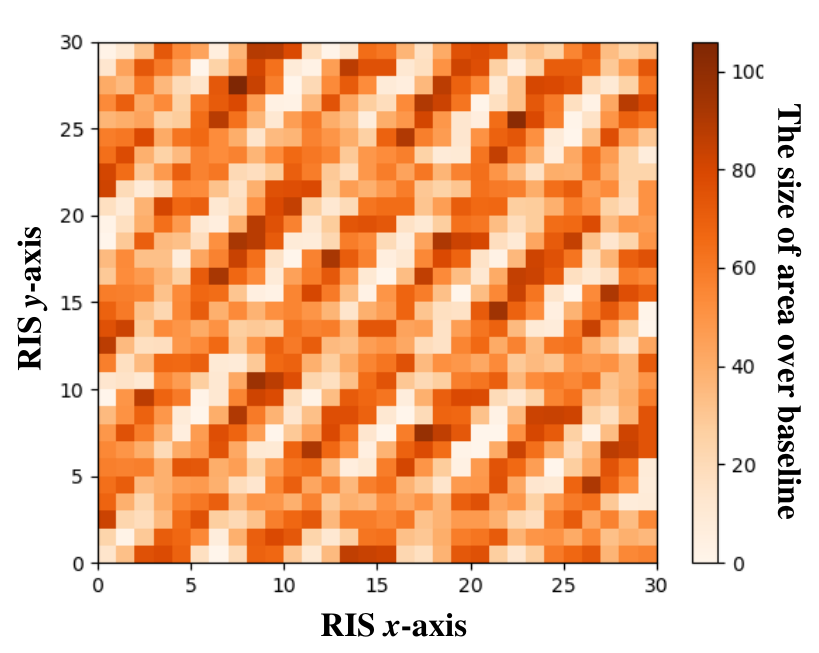}   
      \end{minipage}  
      }
      \caption{The superposition (a) schematic diagram, (b) on (45°,30°) and (45°,60°) directions, and (c) on (30°,40°) and (50°,60°) directions.
      } \label{fig:superposition}
    \end{figure*} 
    
The reflection profile of each direction shows unique periodicity, leading to varying performance with different overlapping methods. A simple superposition of multiple single-reflection profiles for a multi-reflection profile often proves insufficient in performance optimization. It serves multiple beams simultaneously, however causing interference with each specific beam. Therefore, discovering an optimal superposition method to enhance the overall performance of the RIS becomes crucial. Notably, the reflection performance of the RIS should be optimized solely by manipulating the reflection profiles, ruling out the hardware or algorithmic modifications and avoiding additional complexity to the system.

A series of superposition studies of multiple reflective profiles has been conducted to verify the feasibility and performance of the profile configuration. In a single-incidence-two-reflection scenario, we set the single-reflection profile of direction~$2$ four times bigger than the profile of direction~$1$, as shown in Fig.~\ref{fig:schematic}. In this figure, the red and blue rectangles represent the single-reflection profiles of direction~$1$ and direction~$2$ respectively, which are denoted as \textit{profile-$1$} and \textit{profile-$2$} in the following sections. The combination of profile-$1$ and profile-$2$ forms the overlapped multi-reflection profile.
The area enclosed by the yellow dotted line is the moving/searching window, defining the range in which the central point of profile-$2$ can move. The length of this window is determined by the maximum period length of the single reflective sub-profiles. Along with the central point of profile-2 shifts to different positions, new superposition profiles are achieved, referred to as different overlapped positions (or superpositions at different positions) in this paper.

For verification, we move the profile-$2$ and have an exhaustive search with its center point traversing through each cell of the profile-$1$, then calculate the far-field radiation pattern on each position. We set the peak value of the original far-field pattern as the baseline $E_{baseline}$, then calculate the size of the area that has a higher value over the $E_{baseline}$ within the circle centering at the given reflect directions with $20$ degrees radius for each movement. The results is denoted as $A_{m}$ and its indication is visualized in Fig.~\ref{fig:original_2} and the corresponding math equations are given below: 
\begin{equation}
\begin{split}
    & \hspace{40pt}A_{m}=\iint_{D_{m}}f_{m}(\theta_f, \varphi_f)d\sigma
    \\
    & \text{s.t.} \hspace{6pt}D_{m}:(\theta_f-\theta_m)^{2}+(\varphi_f-\varphi_m)^{2}\leq R^{2}\,,\\
    & \hspace{20pt}f_{m}(\theta_f, \varphi_f)=\left\{
    \begin{aligned}
    1, \hspace{5pt}E_{(\theta_f, \varphi_f)}\geq E_{baseline}\,;\\
    0, \hspace{5pt}E_{(\theta_f, \varphi_f)}< E_{baseline}\,.\\
    \end{aligned}
    \right.
\end{split}
    \label{eq:continuous_1}
\end{equation}
where $D_m$ indicates the circle area centering at $(\theta_m, \varphi_m)$ for the \textit{m}-th expected reflect direction with radius \textit{R}~($20$~degrees here), $d\sigma$ is the integral element, and $E_{(\theta_f, \varphi_f)}$ is the far-field value at $(\theta_f, \varphi_f)$. This calculation reflects how much energy can be converged to the reflective direction expected and the sum of these values is the total radiation performance of this RIS reflectarray. The schematic diagram and results are recorded in Fig.~\ref{fig:superposition}. From Fig.~\ref{fig:exhausted_1} and  Fig.~\ref{fig:exhausted_2}, it can be seen that:
\begin{itemize}
\item The superposition of different positions significantly impacts the far-field pattern of the RIS. By adjusting the position of each sub-profile, the far-field performance can be improved.

\item The performance distribution of superposition at different positions shows significant periodicity. Hence, the maximum value can be found within a small scope of distribution by leveraging this periodicity, obviating the need for exhaustive searching.

\item The transition from a poor position to the best position shows a gradual gradient. Hence, the problem can be reformulated as finding the shortest path from any random initial position to the optimum, which can be represented as a typical RL problem.
\end{itemize}

\begin{figure}[t]   
        \centering   
        \includegraphics[width=1\columnwidth]{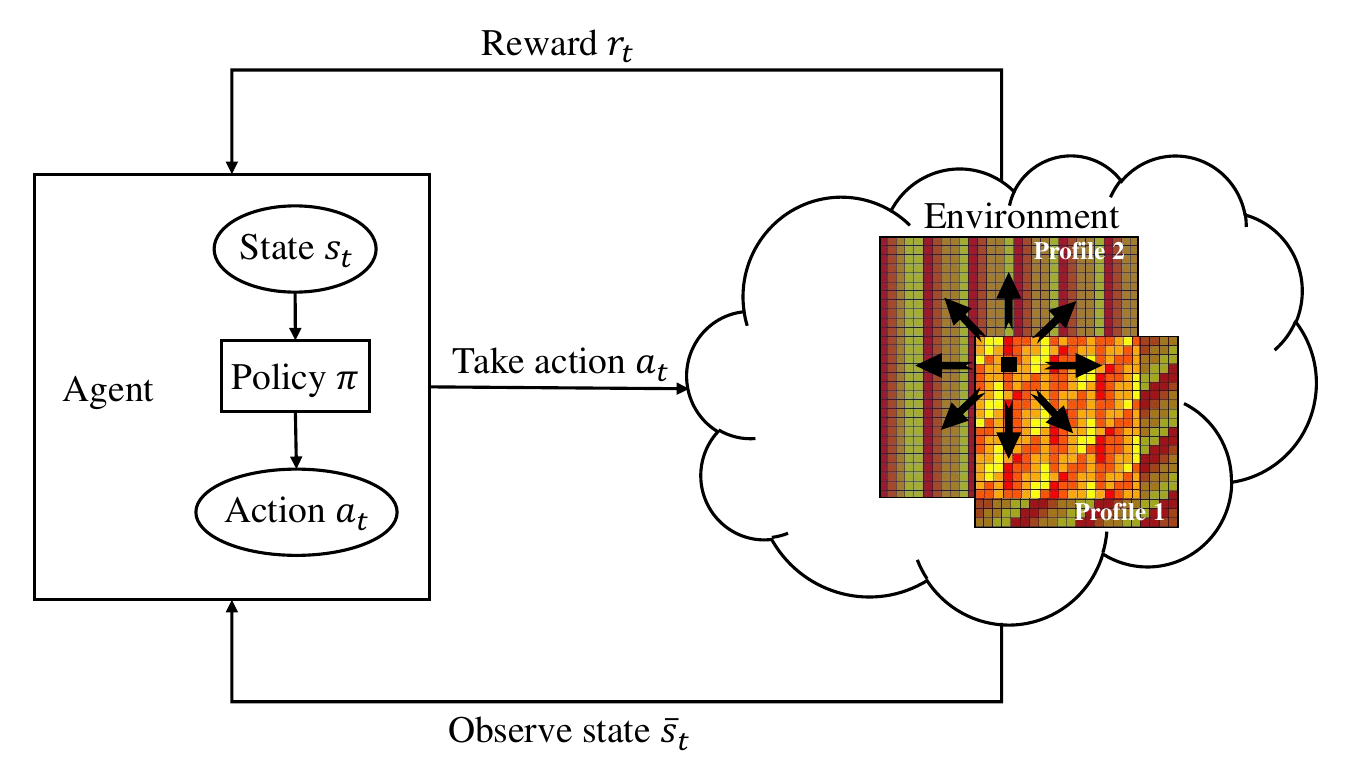}   
            \caption{System diagram of DRL framework} \label{fig:framework}
\vspace{-0.2cm}
\end{figure} 

Based on the simulations and analysis presented above, the combination of RIS and DRL exhibits an intrinsic value to enhance the reflection of RIS-assisted multiple-receiver communication. The optimization objective $A_{total}$ is defined as:
\begin{equation}
    A_{total}=\rm{max}(\sum_{\textit{m}=1}^{\textit{M}}\textit{A}_{\textit{m}})\,,
\label{eq:objective}
\end{equation}
where \textit{M} is the number of reflective directions this RIS reflectarray should achieve and the sum of $A_m$ for different directions-\textit{m} is the total radiating performance of multiple reflect-directions.

\section{DRL-Based Solution}
\label{sec:RL}
\subsection{Deep Q-Networks}
Deep Q-networks (DQN) is a relatively mature and widely used model of RL. The core idea of DQN is using the neural network $f_\theta$ to estimate the state ($s$)-action ($a$) value ($Q$ value). 
Taking the action $a$ in the given space, the optimal policy can be constructed as \cite{liindoor}:
\begin{equation}
    \pi^*(s) = \arg \max Q^*(s,a)\,.
\end{equation}
\eqr{Q^*(s, a)} obeys the Bellman optimality equation \cite{sutton2018reinforcement}:
\begin{equation}
\label{eq:bellman optimality}
    Q^\pi(s,a) = \mathds{E}_{s^\prime \sim \mathcal{S}} \Big[ r + \gamma \max_{a^\prime} Q^* (s^\prime, a^\prime) \big| s,a\Big]\,.
\end{equation}
To learn the $Q$ value at iteration \eqr{i}, the following loss is minimised with respect to \eqr{\theta}:
\begin{equation}
    L_i (\theta_i) = \mathds{E}_{s,a \sim \rho(\cdot)} \Big[ (y_i - Q(s,a;\theta_i))^2 \Big]\,,
\end{equation}
where
\begin{equation}
    y_{i, Q} = \mathds{E}_{s^\prime \sim \mathcal{S}} \Big[ r + \gamma \max_{a^\prime} Q(s^\prime, a^\prime; \theta_{i-1} \big| s,a) \Big]\,.
\end{equation}

\subsection{Markov Decision Process Formalization}
\label{subsec:MDP}
We formulate the problem in (\ref{eq:objective}) as a finite Markov decision process (MDP).
An MDP is defined by the tuple $(\mathcal{S}, \mathcal{A}, \mathcal{P}, \mathcal{R}, \gamma)$, where the set of environment states is represented by $\mathcal{S}$; The action space is denoted by $\mathcal{A}$; $\mathcal{P}$ is the transition probability from state $s\in \mathcal{S}$ to state $s^\prime \in \mathcal{S}$ for any given action $a \in \mathcal{A}$, and $\mathcal{R}$ is the reward function. $\mathcal{\gamma}$ is the discount factor. The DQN agent aims to interact with the environment in order to achieve the largest accumulative rewards. We give the definitions of \textit{state}, \textit{action}, \textit{reward} as below:

\subsubsection{State}
The initial state is an essential one-channel image with the size of $30\times30$ as shown in Fig.~\ref{fig:original_1}. For each pixel of this $30\times30$ array, the color comes from a four-color set which indicates the four different quantization phase responses that the 2-bit RIS element can achieve. The rows and columns of this array represent the positions of each RIS element. Considering the low complexity of the image (the image consists of periodical elements/blocks), we convert this image to a $1\times900$ vector. This vector is used as the state $s_t$ in this work, fed into the fully connected neural network for further processing.

\subsubsection{Action}
The action $a_t$ is a series of discrete values in the range of $[0,8]$ which indicate the eight moving directions plus one standing down action as shown in Fig.~\ref{fig:framework}. The profile-2 will move one cell at each step to its adjacent cell along the moving direction to achieve a different superposition profile for multiple reflect directions.

\subsubsection{Reward}
Based on the analysis in subsection \ref{subsec:Problem Formulation}, the result of exhaustive search shows that the optimization problem of RIS can be transferred to finding the shortest path from any random initial position to the best position. Also because of the gradual gradient in the moving, the reward $r_t$ can be simply set as the size of the area with the far-field value higher than the baseline which is the highest value in expected directions in the original run. The continuous range with $\theta_f\in [0^{\circ},90^{\circ}]$ and $\varphi_f\in [0^{\circ},180^{\circ}]$ can be converted to a discrete range and the size of area can be expressed as the sum of the cells meet the requirements. The equation~(\ref{eq:continuous_1}) can be rewrite as:

\begin{equation}
    \begin{split}
    & \hspace{40pt}r_t=\sum_{m}\sum_{d\in D_m}f_{m,d} \\
    & \text{s.t.} \hspace{6pt}D_{m}:(\theta_f-\theta_m)^{2}+(\varphi_f-\varphi_m)^{2}\leq R^{2}\,,\\
    & \hspace{20pt}f_{m,d}=\left\{
    \begin{aligned}
    1, \hspace{5pt}E_{(\theta_f, \varphi_f)}\geq E_{baseline}\,;\\
    0, \hspace{5pt}E_{(\theta_f, \varphi_f)}< E_{baseline}\,.\\
    \end{aligned}
    \right.
\end{split}
\end{equation}
where $m$ is the number of reflect directions and $m=1,2$ here, $d$ is the pixel within $D_m$.

\subsubsection{Hyperparameters}
The hyperparameters of DQN used in this work can be found in Table \ref{table:Hyperparameters}. The hyperparameters in this table are reasonable empirical values, based on our experience of running many simulations with varying values.

\begin{table}[t]
\centering
\caption{Hyperparameters of DRL}
\label{table:Hyperparameters}
\begin{tabular}{ll} % {l|l}
\toprule
Name & Value\\ \midrule
RL algorithm & DQN \\ 
Exploration rate $\epsilon$ & 0.9\\ 
Batch size & 128\\ 
Maximum time-step in each episode & 11 \\ 
Target network update interval & 100 \\ 
Reward discount factor$\gamma$ & 0.98 \\ 
Optimizer & Adam \\ 
Learning rate & 0.001 \\ 
NN type & Fully connected network \\ 
Number of neurons of each layer & [1000,500,100,50] \\ 
Activate function (not for output layer) & Relu \\ 
Activate function for output layer & Linear \\ 
\bottomrule
\end{tabular}
\end{table}

\section{Numerical Results}
\label{sec:numerical results}
\subsection{Baseline 1: Random Search}
\label{subsec:baseline 1}
The maximum timestep in each episode, as shown in Table~\ref{table:Hyperparameters}, is set to half of the diagonal length of the moving window, which is estimated from the single period of the overlapped reflective profile. This maximum number of timesteps is also used for the random search without DRL, allowing us to compare the performance with the proposed DRL method under the same conditions. In random search, the central position of profile-2 $(\theta_2, \varphi_2)$ is started from the same position as the DRL method but the moving direction of each step is generated randomly. After a series of steps no more than maximum timestep, the random search can also achieve a group of results under different superpositions.

\subsection{Baseline 2: Exhaustive Search}
\label{subsec:baseline 2}
Another baseline is established using an exhaustive search approach. When moving the position of profile-2 to achieve superposition on profile-1, the size of the far-field results is significant but constrained. Although employing an exhaustive search can identify the best position, as depicted in Fig.~\ref{fig:superposition}, it may come at the cost of increased computation time.
\subsection{DRL Model Evaluation and Comparison}
\label{subsec:compare}

% \begin{figure}[t]   
%     \subfloat[\label{fig:training_curve}]{
%       \begin{minipage}[t]{0.45\linewidth}
%         \centering 
%         \includegraphics[width=1.75in]{curve.pdf}   
%       \end{minipage}%
%       }
%       \hfill
%         \subfloat[\label{fig:performance}]{
%       \begin{minipage}[t]{0.55\linewidth}   
%         \centering   
%         \includegraphics[width=1.75in]{performance.pdf}   
%       \end{minipage} 
%       }
%       \caption{(a) The training curve of DQN; (b) The performance of each step in different methods, the DRL validations are based on the 2-bit resolution.
%       }
% \vspace{-0.5cm}
% \end{figure} 
\begin{figure}[t]   
    \subfloat[\label{fig:training_curve}]{
      \begin{minipage}[t]{1\linewidth}
        \centering 
        \includegraphics[width=2.45in]{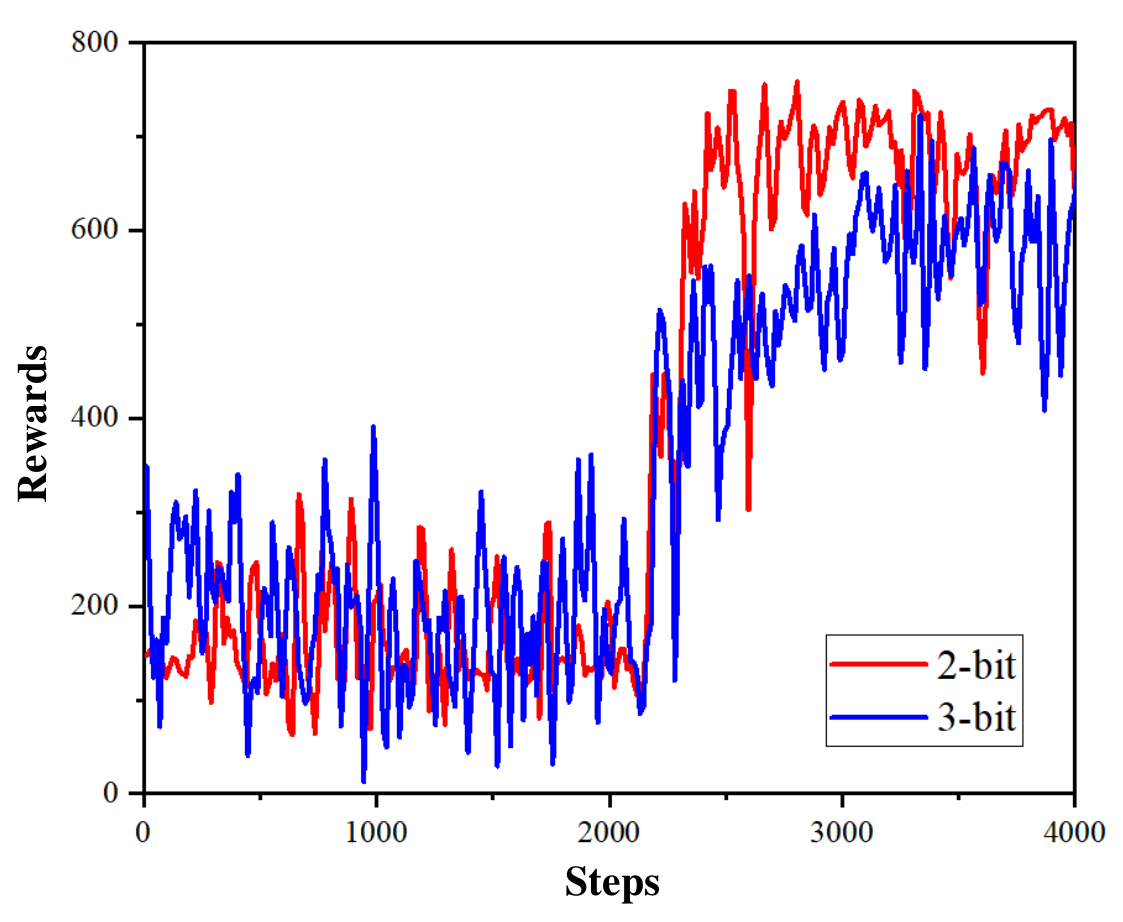}   
      \end{minipage}%
      }
      \vspace{-0.3cm}
      \hfill
        \subfloat[\label{fig:performance}]{
      \begin{minipage}[t]{1\linewidth}   
        \centering   
        \includegraphics[width=2.5in]{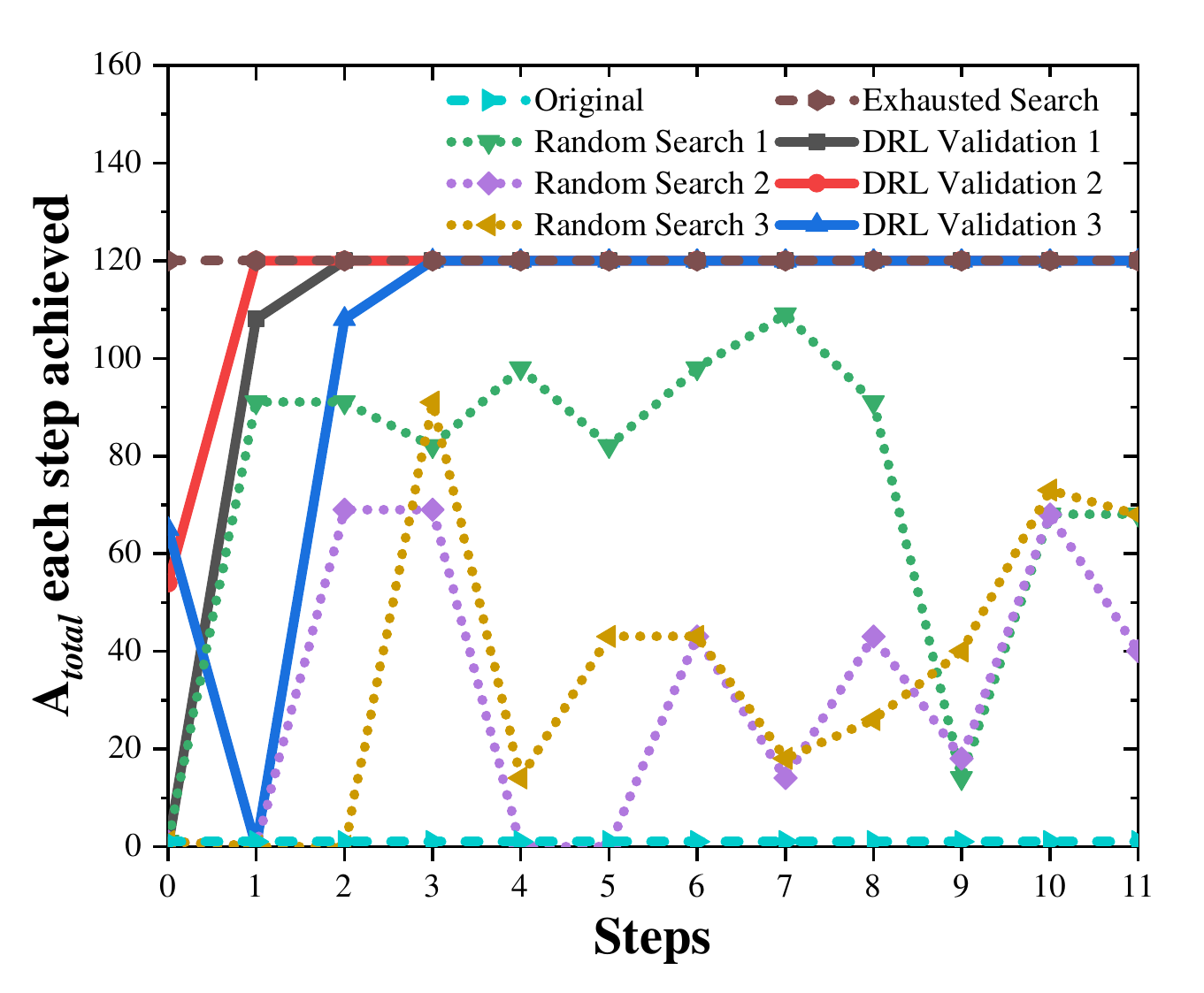}   
      \end{minipage} 
      }
      \caption{(a) The training curve of DQN; (b) The performance of each step in different methods, the DRL validations are based on the 2-bit resolution.
      }
\vspace{-0.5cm}
\end{figure} 

% \begin{figure}[t]   
%         \centering   
%         \includegraphics[width=0.78\columnwidth]{curve.pdf}   
%             \caption{The training curve of DQN} \label{fig:training_curve}
% \end{figure}
% \begin{figure}[t]   
%         \centering   
%         \includegraphics[width=0.8\columnwidth]{performance.pdf}   
%             \caption{The performance of each step in different methods, the DRL validations are based on the 2-bit resolution.} \label{fig:performance}
% \end{figure}
\begin{figure}[t]   
    \subfloat[\label{fig:comparison_1}]{
      \begin{minipage}[t]{0.45\linewidth}
        \centering 
        \includegraphics[width=1.7in]{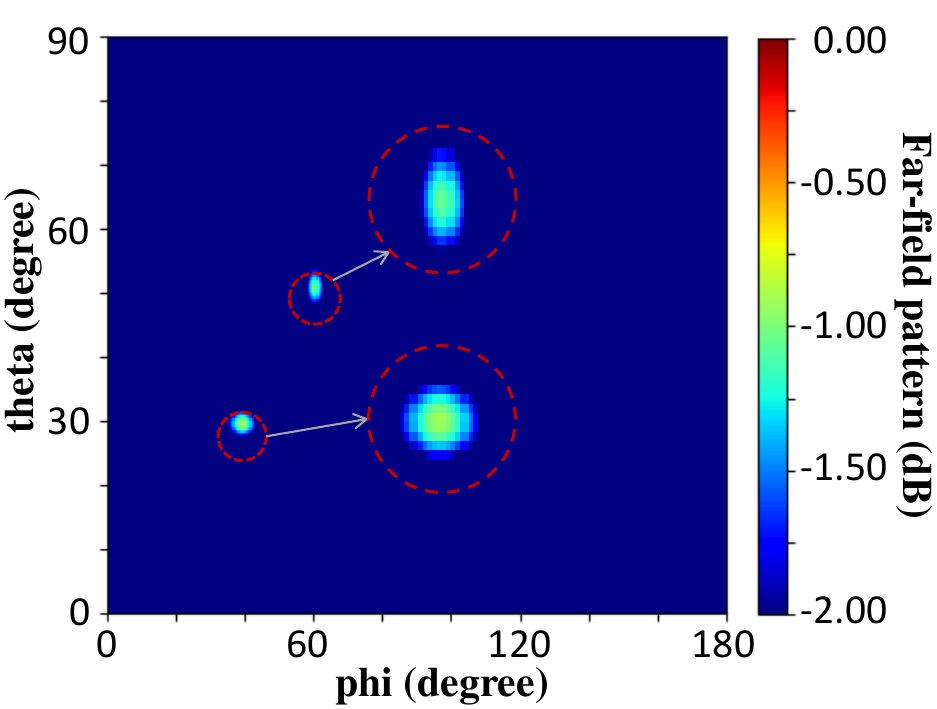}   
      \end{minipage}%
      }
      \hfill
        \subfloat[\label{fig:comparison_2}]{
      \begin{minipage}[t]{0.55\linewidth}   
        \centering   
        \includegraphics[width=1.7in]{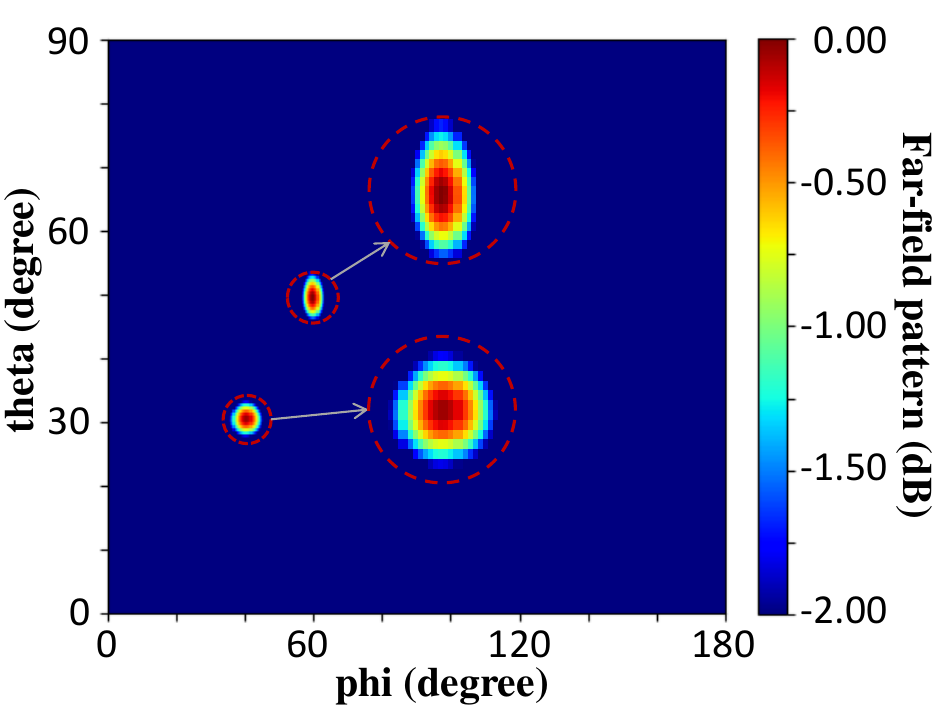}   
      \end{minipage} 
      }
      \caption{The comparison of far-field patterns of the 2-bit resolution (a) before and (b) after DQN adoption.
      } \label{fig:comparison}
\vspace{-0.2cm}
\end{figure} 
In the designed DRL environment, the size of the searching window is $18\times 10$ and the maximum time-step is 11. In the DRL training process, for each episode, the position of the profile-2 will move to the adjacent cell from step to step without exceeding the window range and the maximum time step. The direction of each movement is generated by the exploration and exploitation scheme in the designed DQN network as shown in Table~\ref{table:Hyperparameters}.

We intended to invalidate our proposed method based on different resolutions. However, the single-reflection profile with 1-bit resolution always has a mirrored reflection beam with the same intensity as the intended beam, let alone the multi-reflection profile overlapped from several single-reflection profiles. It will make a great baneful influence on RIS's far-field performance. Hence, Fig.~\ref{fig:training_curve} only shows the smoothed training curve of 2-bit and 3-bit resolution, where the y-axis is the reward that each episode can achieve and the x-axis is the running steps. Comparing these two resolutions, the 3-bit resolution has a little lower increment than the 2-bit resolution which is because a higher resolution is closer to the continuous phase shift with better original performance before DQN training. Meanwhile, the proposed method can also be extended to more than 3-bit resolution.

After that, the trained DQN agent can be exported to validate the performance, compared with the original profile without any move, random search, and exhaustive search is drawn in Fig.~\ref{fig:performance}. For preciseness, the maximum time-step of the random search is also set to 11 steps which is equal to that in the DRL. The random search was run three times from the same starting point while the DRL validation was also run three times from different but adjacent starting points. The result shows that after several steps, the DRL method can make the superposition reach the position with the optimal performance quickly. In contrast, the random search needs a long period of running and then may go through a good position but cannot stand there. Also, it cannot achieve the best position in most runs especially when the optimal position is sparse in the entire search area.

The comparison of far-field patterns of the 2-bit resolution before and after DQN adoption is shown in Fig.~\ref{fig:comparison}, where $1.2$ dB improvement of the peak value and a wider beam have been achieved. Table~\ref{table:Comparison} lists the accuracy, max $A_{total}$ achieved, and time cost of each method. Compared with the exhaustive search, the DRL method can reach the position with the max $A_{total}$ within the shortest time in the same size as the searching window. Though exhaustive search can always find the best value, the DRL method can also find it if the learning process is completed.

\begin{table}[t]
\centering
\caption{Comparison of Different Methods}
\label{table:Comparison}
\begin{tabular}{llll} % {l|l}
\toprule
Name & Random & exhaustive & DRL(proposed)\\ \midrule
Accuracy & \cellcolor{pink}low & \cellcolor{green}highest & \cellcolor{lime}high\\
$A_{total}$ achieved & \cellcolor{pink}\hspace{2pt}/ & \cellcolor{green}120 & \cellcolor{green}120\\
Time cost (\textit{s}) & \cellcolor{lime}36.16 & \cellcolor{pink}504.01 & \cellcolor{green}11.70\\
\bottomrule
\end{tabular}
\end{table}

% \section{Discussions}
% \label{sec:discussion}
% We study the optimal superposition locations for a multiple-reflecting RIS system, aiming to reach its potential maximum performance, and we combine the DRL with it for the first time. There is still much work needed to be done in the future.
% The distribution of far-field peak value is sparse and if the maximum value is not included in the searching window utilized, it will exist an improving space though the max-value in the searching window still can be found. The DRL learning and its result is also restricted to the limited reflect directions simulated. To increase the generalization ability of the trained model, more directions should be considered and a more general policy function should be achieved which can be used for finding the optimal performance for any given reflect directions quickly. To achieve this target, more advanced DRL algorithms will be studied in the future and more delicate MDP formulations should be considered.

\section{Conclusions}
\label{sub:conclusions}
In this work, a DRL-based method to optimize the far-field radiation pattern of multiple-receiver RIS reflectarray has been proposed. According to the simulation results, a $1.2$ dB improvement in the peak gain and a wider beam has been realized. Notably, the accuracy of DRL surpasses that of random search, while its time efficiency greatly outperforms exhaustive search. These findings underscore the suitability of the DRL-based optimization approach in elevating the radiation performance of RIS reflectarrays. Moreover, the DRL-based method proves its efficiency and potential in array synthesis of reconfigurable reflectarrays.
Based on the conclusions above, we will attempt to introduce the DRL in solving more complicated RIS applications. In our future work, we will utilize the DRL-based method to suppress the elevated sidelobe levels of multi-reflect RIS that badly impact energy efficiency and co-channel interference.

\section*{Acknowledge}
\label{sub:ack}
This work was developed within the Innovate UK/CELTIC-NEXT European collaborative research and development project on AIMM (AI-enabled Massive MIMO), and was partially supported by China Scholarship Council for funding the author under No.202108060224.

% \begin{figure}[t]   
%         \centering   
%         \includegraphics[width=0.8\columnwidth]{system_figure.pdf}   
%             \caption{System diagram of FRL framework} \label{fig:system figure}
% \end{figure} 

% To simplify the model, the scenario of two UEs on different directions is considered and simulated in , but the result can also be extended to the situations with more than two UEs. 

\bibliographystyle{IEEEtran} %
\balance
\bibliography{IEEEabrv,references} 

\end{document}